\documentclass[aps,prl,twocolumn,groupedaddress,reprint,showpacs]{revtex4-2}
\usepackage{graphicx}
\usepackage{setspace}
\usepackage{amsmath}
\usepackage{amssymb}
\usepackage{color}
\usepackage{array}
\usepackage{subfigure}
\usepackage{hyperref}
\usepackage{float}
\usepackage{lipsum}

\usepackage[all]{xy}
\newcommand*{\Scale}[2][4]{\scalebox{#1}{$#2$}}%
\newcommand{\RN}[1]{%
  \textup{\uppercase\expandafter{\romannumeral#1}}%
}

\begin{document}
\title{Extended two-body Rydberg blockade interaction with off-resonant modulated driving}
\author{Yuan Sun}
\email[email: ]{yuansun@siom.ac.cn}
\affiliation{CAS Key Laboratory of Quantum Optics and Center of Cold Atom Physics, Shanghai Institute of Optics and Fine Mechanics, Chinese Academy of Sciences, Shanghai 201800, China}

\begin{abstract}
Connectivity has an essential and indispensable role in the cold atom qubit platform. Whilst the two-qubit Rydberg blockade gate recently receives rapid progress on the fidelity side, a pressing challenge is to improve the connectivity in pursuit of genuine scalability, with the ultimate prospect of fully-connected cold atom qubit array. It turns out that a solid step along this direction can be made by introducing extra buffer atom to extend two-qubit Rydberg blockade gate beyond a purely nearest-neighbor two-body interaction. Through Rydberg dipole-dipole interactions, the buffer atom couples with the two qubit atoms which do not directly exert any physical influence on each other. The established method of off-resonant modulated driving is not only convenient but also lays down the groundwork for this latest development. Although the atomic linkage structure here exhibits nontrivial complications compared to previous cases of mere two-body system, the population can satisfyingly return to the ground state after the ground-Rydberg transition with properly designed modulation waveforms. It can be instantiated via one-photon and two-photon ground-Rydberg transitions in common practices. Furthermore, with buffer atom relay or similar structures, it is possible to realize two-qubit entangling gate between two far-away qubit atoms. Besides the core issue that such solutions are attainable, the representative modulation patterns are also analyzed, demonstrating the versatility of buffer-atom-mediated two-qubit gate. Put in a broader perspective, these efforts bring the cold atom qubit platform closer to the notions of wires and junctions in solid state electronics. 
\end{abstract}
\maketitle

The two-qubit entangling gate is a central theme in the development of cold atom qubit platform \cite{RevModPhys.82.2313, RevModPhys.87.1379, J.Phys.B.49.202001}. The importance of inter-atomic dipole-dipole interaction \cite{PhysRevA.62.052302} has repeatedly manifested itself in the theoretical and experimental efforts \cite{PhysRevLett.85.2208, nphys1183, PhysRevLett.104.010503} of two-qubit Rydberg blockade gate research ever since the early days, as well as the fast-evolving fields of quantum simulation and quantum precision measurement with Rydberg atoms \cite{Browaeys2020review, Browaeys2021nature, Lukin2021nature, Kaufman2022nphys, Browaeys2023nature, Endres2023nature}. In particular, as the quality of cold atom qubit array \cite{PhysRevA.92.022336, Weiss2581Science, PhysRevLett.122.203601, Saffman2019prl, Kaufman2020nature, PhysRevApplied.16.034013} and the fidelity of single-qubit gate \cite{PhysRevLett.114.100503, PhysRevLett.121.240501, nature604.457, PhysRevX.12.021027} are making steady progress towards a promising level, the quest for scalability of cold atom qubit platform entails extensive and in-depth work to improve the fidelity and connectivity of two-qubit gates. Theoretical investigations have revealed that special modulation in the atom-light interaction process, including amplitude and frequency modulations, can lead to a family of Rydberg blockade gates \cite{PhysRevApplied.13.024059, OptEx480513} via a smooth pulse inducing conditional phase changes, with potential extensions to ensemble qubits \cite{PhysRevLett.115.093601, PhysRevA.102.042607}. Recently, this intriguing category of two-qubit gates via carefully modulated driving fields has been experimentally demonstrated \cite{PhysRevA.105.042430, Lukin2023nature} with considerable improvement in fidelity.

Nevertheless, these latest experimental progress on two-qubit Rydberg gate fidelity still faces the severe limitations of nearest neighbor interaction to a large extent. The Rydberg dipole-dipole interaction has a far more delicate structure than the idealized Rydberg blockade effect and the strength falls off rapidly with respect to the distance \cite{RevModPhys.82.2313, JPB_Singer_2005, PhysRevApplied.7.064017, nphys1183, PhysRevLett.104.010502, PhysRevLett.119.160502}. Take the scaling of $R^{-6}$ as an example, doubling the qubits' distance $R$ indicates a 64-fold reduction in the Rydberg blockade strength. On the other hand, if two qubit atoms are too close to each other, then the issue of crosstalk \cite{PhysRevA.107.063711} will impede the practical performance of gates. Although the method of Rydberg interaction strength adaption (RISA) has been devised to accommodate finite Rydberg blockade strength \cite{OptEx480513} for the off-resonant modulated driving (ORMD) gate protocols, a systematic approach to upgrade the two-qubit Rydberg blockade gate towards the fully-connected cold atom qubit array still remains elusive so far.
Furthermore, keeping a highly coherent ground-Rydberg transition \cite{nphys3487, PhysRevLett.104.010502, PhysRevApplied.15.054020, PhysRevLett.129.200501, YouLi2023arXiv} is also a crucial factor for the fidelity and connectivity, which implicitly imposes extra subtle requirements about the waveforms of ORMD protocols, including the gross smoothness, non-abrupt turn-on and turn-off stage, suppression higher-order frequency components and so on \cite{Yuan2023arXiv}.   

Facing the challenges of connectivity and fidelity, we will place an extra buffer atom into the system of qubit atoms to form an extended Rydberg blockade interaction and subsequently derive the buffer-atom-mediated (BAM) two-qubit gates. The rest of contents starts with the generalized principles behind ORMD methods about binging forth specific multi-atom state changes with well-controlled atom-atom interactions. Then we move on to introduce the key ingredients of extended Rydberg blockade interaction as well as explain the underlying physical mechanisms. Although the situation of BAM gates seems much more complicated than the purely two-body scenario, our investigations reveal that not only do such category of solutions exist, but also appropriate solutions can be found with reasonable parameters of practical experimental conditions. Eventually, specific examples will be demonstrated and we will discuss the framework of buffer atom relay. Throughout the text, the gate fidelity is evaluated according to Refs. \cite{JAMIOLKOWSKI1972275, CHOI1975285, PhysRevA.71.062310, PEDERSEN200747}.

\begin{figure}[h]
\centering
\includegraphics[width=0.432\textwidth]{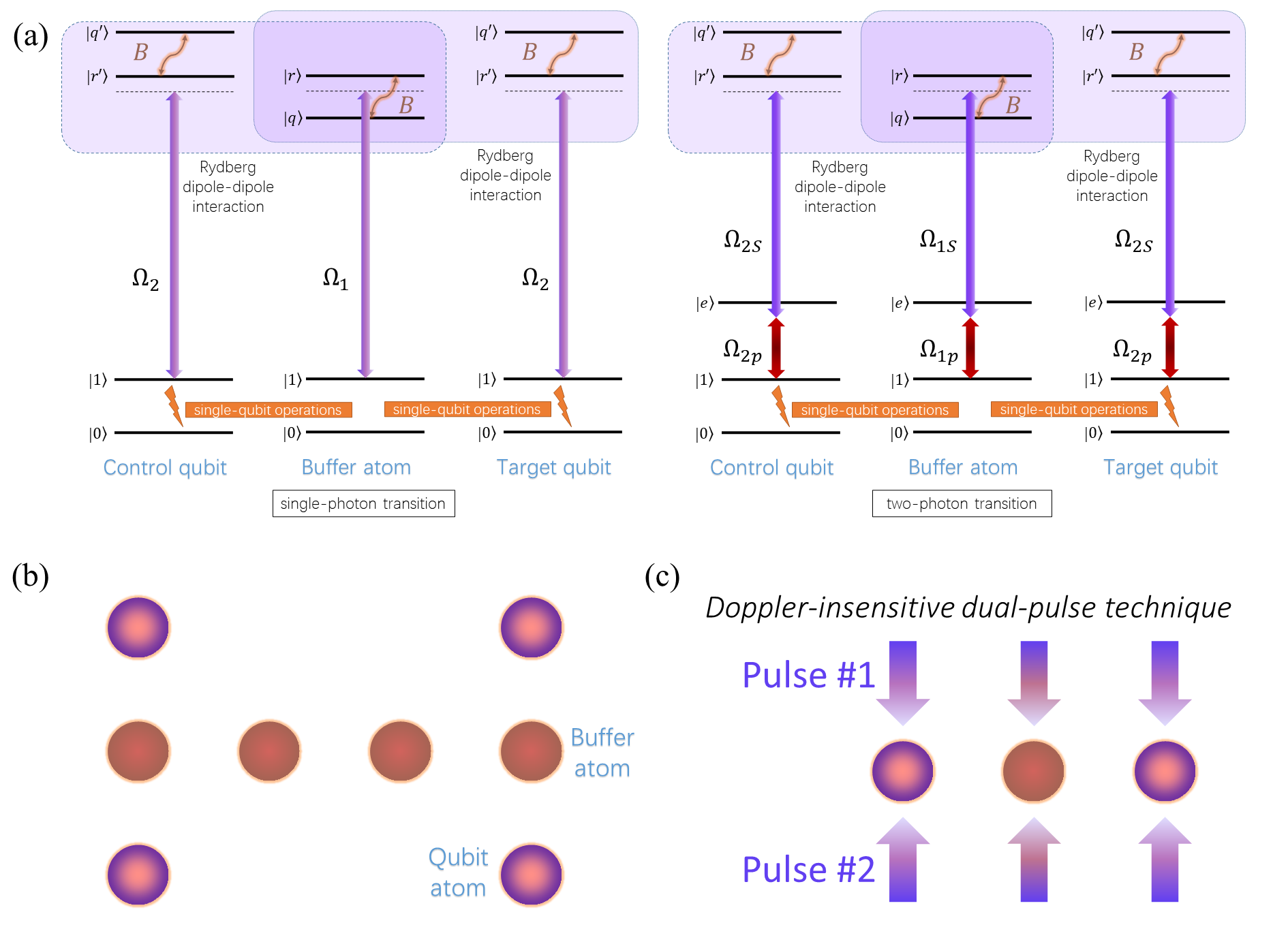}
\caption{(Color online) (a) Atomic levels and transitions for the buffer atom mediated Rydberg blockade gate, where in fact the choice of single-photon or two-photon transition does not have to be the same for all qubit and buffer atoms. (b) Buffer atom relay. (c) Dual-pulse technique for Doppler-insensitive gate operation. Here, we denote the qubit register states as $|0\rangle, |1\rangle$ and the relevant Rydberg state as $|r\rangle, |r'\rangle$.}
\label{fig1:layout_sketch}
\end{figure}

Generally speaking, the principles of ORMD method can be understood as in interacting with a continuous but off-resonant pulse, the qubit atoms' wave functions receive the correct state-dependent conditional phase shifts, where the populations return to the initial qubit register states. Such a special time evolution emerges as a nontrivial result of carefully tailored external optical or microwave driving under the presence of well-calibrated inter-atomic interactions. The induced entanglement originates fundamentally from the state-dependent multi-body interactions between qubit atoms, with the particularly convenient choice of Rydberg dipole-dipole interaction that can be coherently turn on and off. On the other hand, compatibility with lab hardwares and robustness against adverse effects are also priorities of the design process, and the eventual fidelity in realistic experimental implementations sets the ultimate criteria. Therefore, the quantum physics underlying the process requires careful analysis and finding the appropriate modulation styles is of the essence. More specifically, a waveform $w$ for Rabi frequencies and detunings can be expressed with respect to a complete basis $\{g_\nu\}$ for $L^2$ functions on the prescribed time interval: $w=\sum_{\nu=0}^{\infty} \alpha_\nu g_\nu$ with coefficients $\alpha_\nu$'s. Such a framework allows adequate descriptions of possible solutions in the abstract sense, while truncation in the expansion operates conveniently in practice. In particular, the waveforms in this paper are generated on basis of Fourier series, equivalent to the method of suppressing high-frequency components \cite{Yuan2023arXiv} from the mathematical point of view. Additionally, as part of the efforts in reducing abruptness of on-off switch stages, the first-order derivatives at the two end points are zero, resulting in a `tranquilized' waveform.

Fig. \ref{fig1:layout_sketch}(a) shows the basic ideas of a buffer atom participating in the two-qubit system. While the distance between the two qubit atoms already utterly reduces the Rydberg dipole-dipole interaction strengths, appropriate choices of the Rydberg states and tuning the atomic resonances with external fields can lead to further suppressing of mutual influence. Therefore, it makes sense to assume that the significant Rydberg dipole-dipole interactions occur only between qubit-buffer atom pairs, while neglecting the Rydberg dipole-dipole interaction of two relatively faraway qubit atoms for now. Technically this assumption equivalently amounts to the approximation of nearest neighbor interaction in such a three-body system. The two qubit atoms' distances to the buffer atom do not need be the same while the symmetric configuration here is for conciseness. We also postulate that the Rydberg blockade shift of the three-body Rydberg excited state $|r'rr'\rangle$ compared to the two-body Rydberg blockade never populated throughout the interaction. In other words, this system consists of extended two-body Rydberg blockade interaction with off-resonant modulated driving, whose details are embedded in the Rabi frequency and detuning terms' waveforms.

\begin{figure}[h]
\centering
\includegraphics[width=0.432\textwidth]{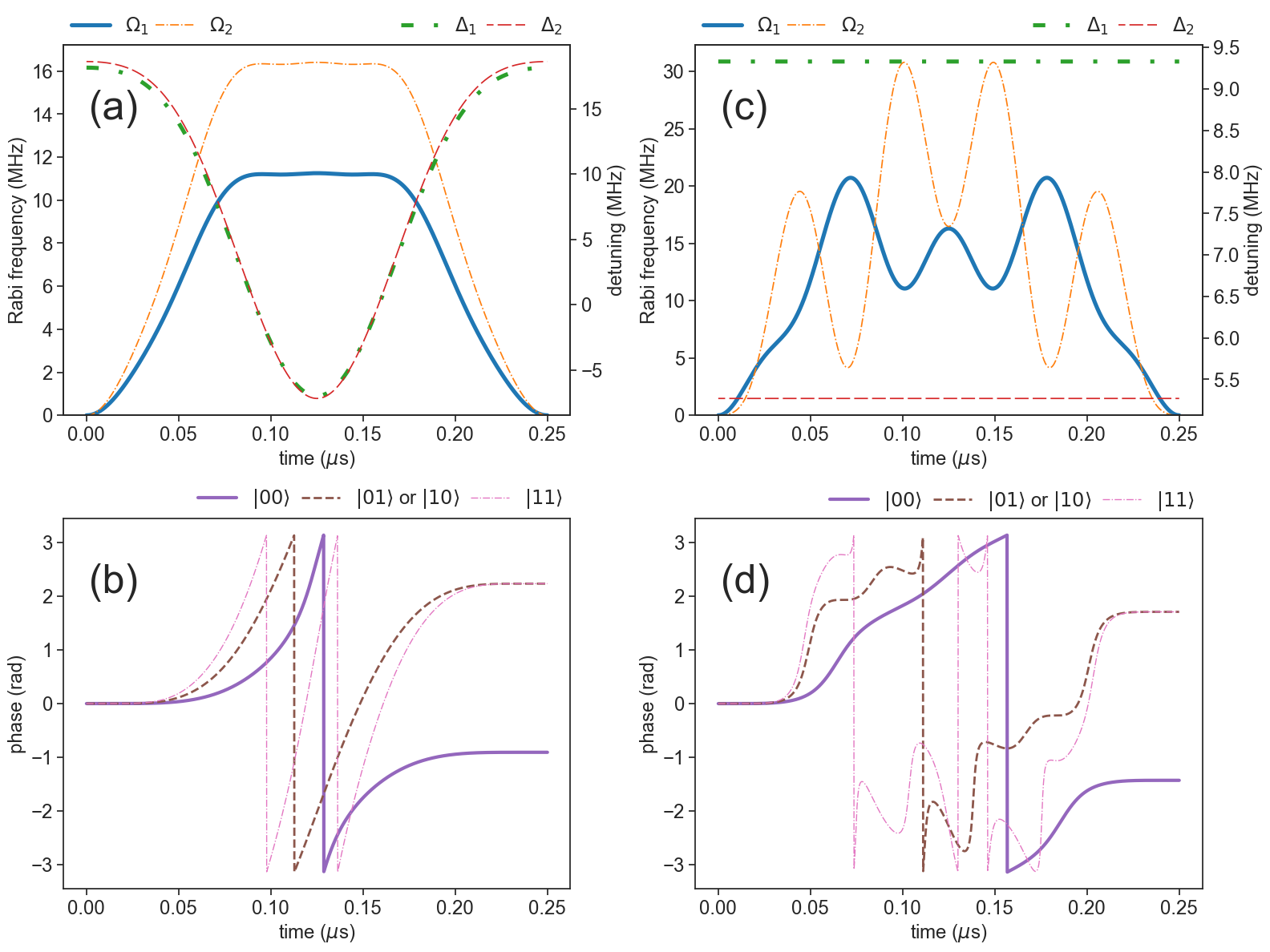}
\caption{(Color online) Sample waveforms of BAM gate with one buffer atom. (a) Hybrid modulation.  (b) Phases of wave functions corresponding to (a). (c) Only amplitude modulation.  (d) Phases of wave functions corresponding to (c). The calculated gate errors are less than $10^{-4}$.}
\label{fig:BAMgate_1photon}
\end{figure}

With the above context and hypothesis, calculation of waveforms becomes possible for the two-qubit entangling gate. The buffer atom is always prepared in the state of $|1\rangle$ before the interaction. Compared with the purely two-body case of ORMD gates, although the levels and linkages here seem significantly more complicated, the key question of finding rightful conditional phase shifts remains the same. More specifically, express the three-body state in the order of control qubit, buffer, and target qubit. The two-qubit state $|0_c0_t\rangle$ corresponds to the three-body state $|010\rangle$, whose dynamics emerges as the ground-Rydberg transition of buffer atom. The two-qubit state $|0_c1_t\rangle, |1_c0_t\rangle$ correspond to the three-body state $|011\rangle, |110\rangle$ respectively, and the two-qubit state $|1_c1_t\rangle$ corresponds to the three-body state $|111\rangle$, whose dynamics involves the Rydberg dipole-dipole interaction between buffer and qubit atoms. We seek the conditional phase shifts on the two-qubit states that satisfy a Controlled-PHASE gate, and in particular Controlled-Z (CZ) gate for the scope of this paper. If the buffer atom becomes a normal qubit atom like the other two, then this routine can produce waveforms of a Toffoli gate in the prescribed system. 

Enticingly, practical solutions adhering to the above concepts exist, opening the door to explore the wide possibilities of producing high-fidelity state changes via highly coherent quantum control process in complicated systems. As previously stated, the routine of representing waveforms by prescribed basis functions converts the abstract notions of continuous functions to a series of coefficients, with which the numerical methods can efficiently operate. Throughout the text, the function $f$ of Rabi frequency or detuning term's waveforms will be expressed in terms of $[a_0, a_1, \ldots, a_N]$ with the interpretation of $f(t)=2\pi \times \big(a_0 + \sum_{n=1}^{N} a_n\exp(2\pi i nt/\tau) + a^*_n\exp(-2\pi i nt/\tau_r) \big)/(2N+1) \text{ MHz}$ for a given reference time $\tau_r = 0.25\, \mu\text{s}$, as explained earlier that the truncated Fourier series assumes the essential role. A sample result of one-photon ground-Rydberg transition is presented in Fig. \ref{fig:BAMgate_1photon} with Rydberg Blockade strength $B=2\pi\times 50 \text{ MHz}$. The commonly available experimental technique of single-site addressing allows each atom to have different Rabi frequency and detuning terms. While the amplitude modulation is often deemed as essential for applying a finite pulse, whether applying the additional frequency modulation depends on the situation and user's choice. For Fig. \ref{fig:BAMgate_1photon}(a), $\Omega_1(t) = 0.686\times\Omega_2(t)$, $\Omega_2(t)$ is given by $[112.83, -46.32, -11.51, 2.35, 0.193, -1.14]$, $\Delta_1(t)$ is given by $[40.14, 31.41, -6.14]$ and $\Delta_2(t)$ is given by $[41.67, 32.23, -6.56]$. For Fig. \ref{fig:BAMgate_1photon}(c), $\Omega_1(t), \Omega_2(t)$ are expressed by $[124.49, -34.38, -28.36, 1.50, 10.93, -11.93]$,  $[153.58, -51.60, 2.09, -23.86, -33.57, 30.15]$ respectively, $\Delta_1=2\pi\times9.33\text{ MHz}$ and $\Delta_2=2\pi\times5.27 \text{ MHz}$. Of course, the solutions are not unique and the specialized suitable waveforms can be calculated according to the specific experimental needs and constraints.

\begin{figure}[h]
\centering
\includegraphics[width=0.432\textwidth]{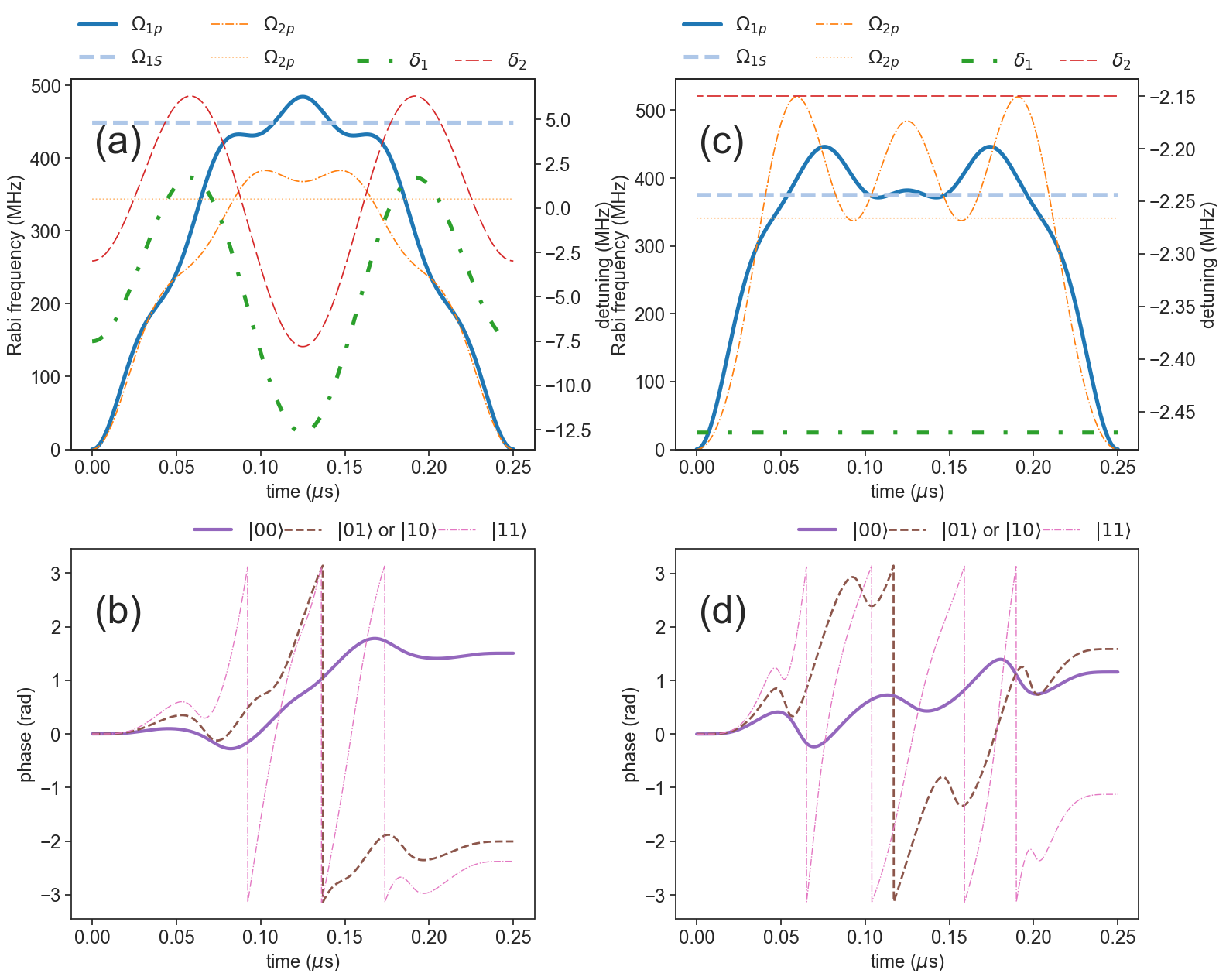}
\caption{(Color online) For two-photon ground-Rydberg transition. (a) Hybrid modulation of $\Omega_p$ and $\delta$.  (b) Phases of wave functions corresponding to (a). (c) Only amplitude modulation of $\Omega_p$.  (d) Phases of wave functions corresponding to (c). The calculated gate errors are less than $10^{-4}$.}
\label{fig:BAMgate_2photon}
\end{figure}

On the other hand, the two-photon ground-Rydberg transition are widely adopted which calls for distinctive attention. While designing waveforms to exactly emulate that of one-photon transition can certainly generate reasonable solutions, utilization of the extra degrees of freedom can provide more opportunities and reduce complexities of implementation. In fact, categorization of the two-photon transition's modulation patterns \cite{OptEx480513} indicates that there exist rich underlying physics and a variety of tools for the experiment to choose from, with or without frequency modulation. Here, we focus on the case of modulating $\Omega_p$ for convenience of implementation and demonstrate the sample results in Fig. \ref{fig:BAMgate_2photon}. Without loss of generality, set the one-photon detuning as $\Delta_0=2\pi\times5\text{ GHz}$,. For Fig. \ref{fig:BAMgate_2photon}(a), $\Omega_{1S}=2\pi\times448.87\text{ MHz}, \Omega_{2S}=2\pi\times343.82\text{ MHz}$, and $\Omega_{1p}(t), \Omega_{2p}(t), \delta_1(t), \delta_2(t)$ are given by [3229.71, -1170.45, -239.25, -33.70, -43.47, -127.98], [2713.71, -909.89, -215.59, -106.97, -129.61, 5.20], [-21.25, 6.44, -14.58] and [2.00, 6.04, -14.49] respectively. For Fig. \ref{fig:BAMgate_2photon}(c), $\Omega_{1p}(t), \Omega_{2p}(t)$ are given by [3442.06, -1350.63, -150.75, -22.40, -81.41, -115.84], [3396.69, -885.46, -654.28, -205.05, 262.51, -216.07] respectively, and $\Omega_{1S}=2\pi\times370.18\text{ MHz}, \Omega_{2S}=2\pi\times321.13\text{ MHz}, \delta_1=2\pi\times-4.03\text{ MHz}, \delta_2=2\pi\times-2.00\text{ MHz}$. The discussed method can obviously switch to different values of Rydberg Blockade strength $B$ which can be experimentally measured. After all, the design of waveforms is capable of adapting to even a more complicated realistic Rydberg dipole-dipole interaction structure \cite{OptEx480513} beyond the demonstration of $B=2\pi\times 50 \text{ MHz}$ here.

\begin{figure}[h]
\centering
\includegraphics[width=0.432\textwidth]{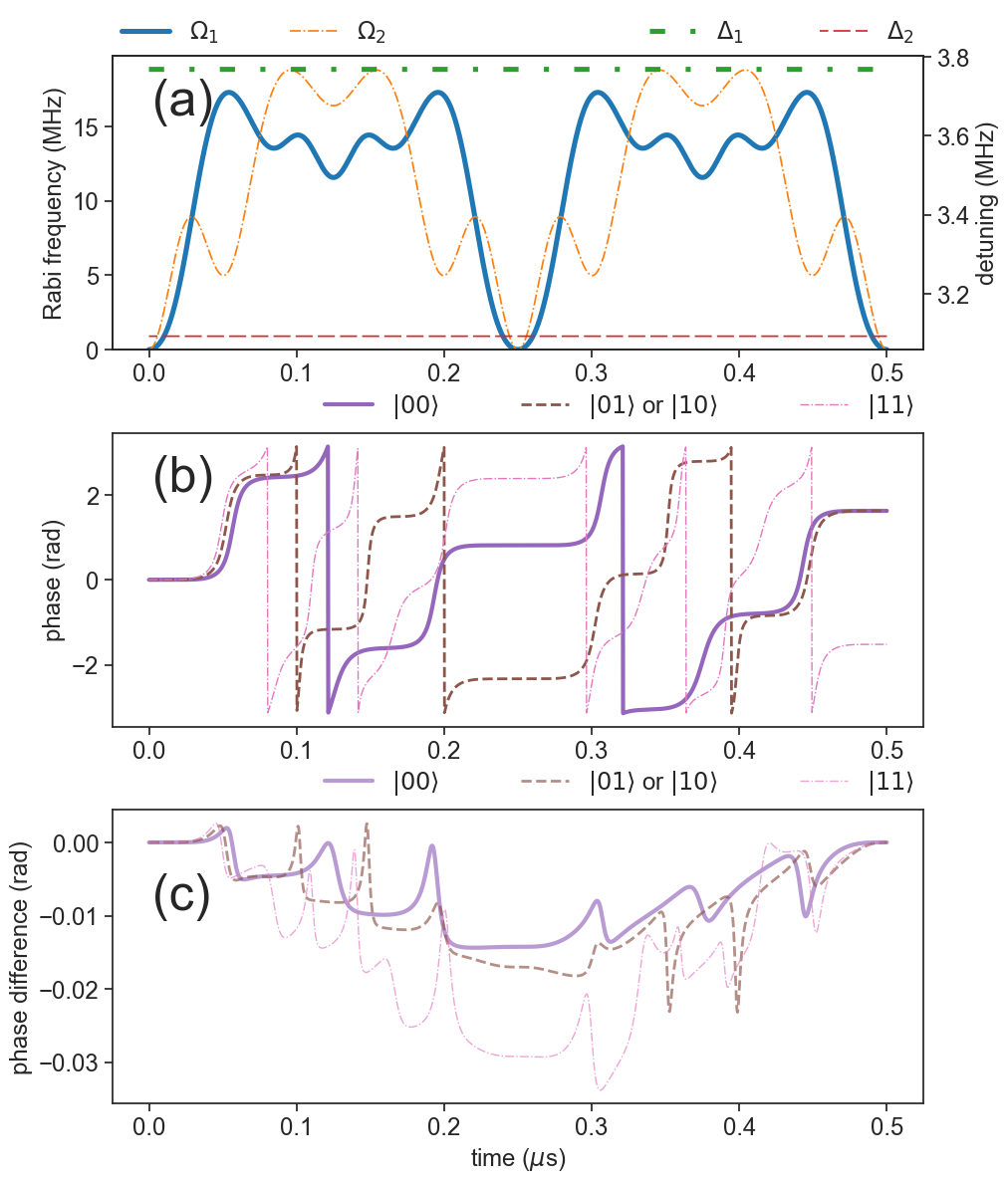}
\caption{(Color online) Doppler-insensitive upgrade via the dual-pulse technique. (a) Waveforms. (b) Phases of wave functions for zero velocity case. (c) Phase differences. The calculated gate error is less than $10^{-4}$.}
\label{fig:dual_pulse}
\end{figure}

The Doppler-induced dephasing caused by the qubit atoms' residual thermal motions \cite{Saffman2019prl, PhysRevApplied.15.054020} creates an inherent challenge to the fidelity. During the interaction with ORMD pulses, this adverse effect usually leads to first-order deviation in phase and second-order deviation in population. The dual-pulse technique \cite{PhysRevApplied.13.024059} has been proposed to tackle with this problem that if two identical driving pulses are applied consecutively but in completely opposite directions, then their first-order effect cancels out, henceforth restoring the fidelity. Fortunately, the framework of BAM gates remains compatible with the dual-pulse technique, as sketched in Fig. \ref{fig1:layout_sketch}(c). Fig. \ref{fig:dual_pulse} shows a typical result, with $\Omega_1(t), \Omega_2(t)$ as [174.55, -33.89, -39.45, -18.46, -3.37, 7.20, -1.07, 1.76],  [164.23, -57.66, 3.13, 6.80, -21.23, -11.14, -2.52, 0.50] respectively, $\Delta_1=2\pi\times3.768\text{ MHz}$ and $\Delta_2=2\pi\times3.093 \text{ MHz}$. Fig. \ref{fig:dual_pulse}(c) shows the phase differences of wave functions between the cases of all the buffer and qubit atoms having nonzero velocities of $kv=25\text{ kHz}$ and zero velocities, and clearly reveals the self-cancellation mechanism of the first-order deviation.

If the control and target qubit atoms are too far away beyond the effective range of one buffer atom, then the buffer atom relay can help to facilitate the two-qubit gate, as illustrated in Fig. \ref{fig1:layout_sketch}(b). For succinctness, we choose subscripts $b, f$ to distinguish the two buffer atoms. Step 1: generate a maximally entangled state of buffer atoms via the CZ gate:
\begin{equation}
|0_b0_f\rangle \,\to\, \frac{1}{2}\big(|0_b0_f\rangle + |0_b1_f\rangle +|1_b0_f\rangle - |1_b1_f\rangle \big).
\end{equation}
Step 2: a CZ gate between buffer atom $b$ and qubit atom $c$ and a CZ gate between buffer atom $f$ and qubit atom $t$, which does not matter if enacted simultaneously or not, as expressed by Table \ref{tab:buffer_atom_2CZ}.
\begin{table}
\begin{center}
\Scale[0.8]{
  \begin{tabular}{ | c | c | }  
    \hline
    state of buffer atoms & state of qubit atoms \\ \hline
    $|0_b0_f\rangle$ & $\big(\alpha_0|0_c\rangle + \alpha_1|1_c\rangle\big)\big(\beta_0|0_t\rangle + \beta_1|1_t\rangle\big)$ \\ \hline
    $|0_b1_f\rangle$ & $\big(\alpha_0|0_c\rangle + \alpha_1|1_c\rangle\big)\big(\beta_0|0_t\rangle - \beta_1|1_t\rangle\big)$ \\ \hline      
    $|1_b0_f\rangle$ & $\big(\alpha_0|0_c\rangle - \alpha_1|1_c\rangle\big)\big(\beta_0|0_t\rangle + \beta_1|1_t\rangle\big)$ \\ \hline  
    $|1_b1_f\rangle$ & $-\big(\alpha_0|0_c\rangle - \alpha_1|1_c\rangle\big)\big(\beta_0|0_t\rangle - \beta_1|1_t\rangle\big)$ \\ \hline    
  \end{tabular}
  }
\end{center}
\caption{Two buffer atoms interact with two qubit atoms in initial state $\big(\alpha_0|0_c\rangle + \alpha_1|1_c\rangle\big)\big(\beta_0|0_t\rangle + \beta_1|1_t\rangle\big)$.}
\label{tab:buffer_atom_2CZ}
\end{table}
Step 3: perform measurement of the buffer atoms as projections onto local basis $(|0\rangle \pm |1\rangle)/\sqrt{2}$, as expressed by Table \ref{tab:buffer_atom_projection}.
Step 4: local phase rotations of qubit atoms if a CZ gate of standard format is asked for. A bootstrap approach can generate the necessary entangled buffer atom pair if buffer atoms $b, f$ are still too far away. That is, use extra buffer atoms $b_1, f_1$ to prompt $b, f$'s entanglement preparation; if $b_1, f_1$ are still too far away, then repeat it until the ends meet. 

\begin{table*}
\begin{center}
  \begin{tabular}{ | c | c | }  
    \hline
    projection of buffer atoms & outcome state of qubit atoms \\ \hline
    $\frac{1}{2}\big(|0_b\rangle + |1_b\rangle\big)\big(|0_f\rangle + |1_f\rangle\big)$ 
    & $\alpha_0\beta_0|0_c0_t\rangle + \alpha_0\beta_1|0_c1_t\rangle + \alpha_1\beta_0|1_c0_t\rangle - \alpha_1\beta_1|1_c1_t\rangle\big)$ \\ \hline
    $\frac{1}{2}\big(|0_b\rangle - |1_b\rangle\big)\big(|0_f\rangle + |1_f\rangle\big)$ 
    & $\alpha_0\beta_0|0_c0_t\rangle - \alpha_0\beta_1|0_c1_t\rangle + \alpha_1\beta_0|1_c0_t\rangle + \alpha_1\beta_1|1_c1_t\rangle\big)$ \\ \hline      
    $\frac{1}{2}\big(|0_b\rangle + |1_b\rangle\big)\big(|0_f\rangle - |1_f\rangle\big)$ 
    & $\alpha_0\beta_0|0_c0_t\rangle + \alpha_0\beta_1|0_c1_t\rangle - \alpha_1\beta_0|1_c0_t\rangle + \alpha_1\beta_1|1_c1_t\rangle\big)$ \\ \hline  
    $\frac{1}{2}\big(|0_b\rangle - |1_b\rangle\big)\big(|0_f\rangle - |1_f\rangle\big)$ 
    & $-\alpha_0\beta_0|0_c0_t\rangle + \alpha_0\beta_1|0_c1_t\rangle + \alpha_1\beta_0|1_c0_t\rangle + \alpha_1\beta_1|1_c1_t\rangle\big)$ \\ \hline    
  \end{tabular}
\end{center}
\caption{State of qubit atoms after measurement of buffer atoms.}
\label{tab:buffer_atom_projection}
\end{table*}

Other methods of instantiating the buffer atom relay exist and we think that some of them can also perform well. For example, consider a system consisting of two qubit atoms and $n$ buffer atoms arranged in 1D array with the two qubit atoms at the two ends. Via localized driving on each atom with carefully calculated waveforms, conditional phase shifts of the two-qubit entangling gate can be acquired with the site-to-site Rydberg dipole-dipole interaction. The states of buffer atoms will remain the same after the manipulations. Such a process will look similar to the main contents of this work just like replacing the one buffer atom with many, which seems as an interesting project for the future.

\begin{figure}[h]
\centering
\includegraphics[width=0.432\textwidth]{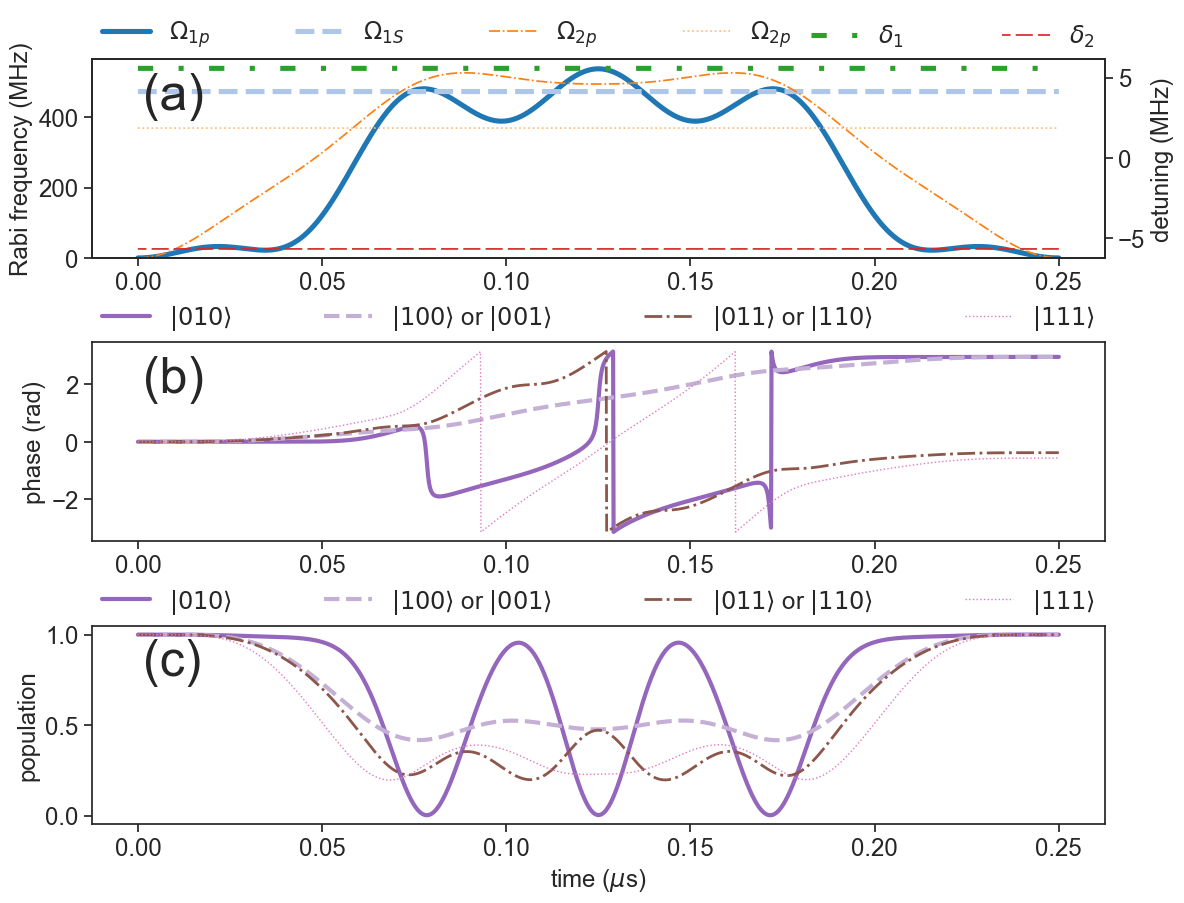}
\caption{Waveforms for the phase gate part of three-qubit Toffoli gate via two-photon ground-Rydberg transition.}
\label{fig:toffoli_2ph}
\end{figure}

Last but not least, if the buffer atom is also perceived as a qubit atom, then a three-qubit Toffoli gate can be instantiated within this extended two-body Rydberg blockade system. Unlike the usual requirement of strong mutual interaction existing for each pair of this three-body system, now the two qubit atoms at the two ends have negligible Rydberg dipole-dipole interaction strength with each other. The key point then comes down to the design of waveforms satisfying the requirement of conditional phase shifts corresponding to the Toffoli gate in one form or another, compatible with this cold atom qubit system. Fig. \ref{fig:toffoli_2ph} show a sample result of three-qubit Toffoli gate via two-photon ground-Rydberg transition, where $\Omega_{1p}(t), \Omega_{2p}(t)$ are given by $[2828.10, -1469.66, -185.18, 285.37, 248.22, -292.80]$, $[3683.49, -1374.63, -434.93, 69.95, -48.91, -53.22]$ respectively, and $\Omega_{1S}=2\pi\times472.49\text{ MHz}, \Omega_{2S}=2\pi\times368.79\text{ MHz}, \delta_1=2\pi\times5.574\text{ MHz}, \delta_2=2\pi\times-5.663\text{ MHz}$.

There exist possibilities for a few potential refinements in the future. The process itself contains rich many-body dynamics and interesting quantum physics, whilst many delicate effects can be later evaluated in greater details such as the residual Rydberg dipole-dipole interactions between the two qubit atoms, the three-body Rydberg dipole-dipole interactions when both qubit and buffer atoms are excited, the case of buffer atom chains embedded with strong Rydberg blockade effect between neighbors and so on. The subtle dissipation effects embedded in the Rydberg dipole-dipole interactions also merit further careful investigations. Extra subtle dissipation processes during the Rydberg dipole-dipole interaction processes also call for further investigations. More refinements are expected about better quality of dark state mechanism, less non-adiabatic effects \cite{PhysRevLett.103.110501, PhysRevA.90.033408} and smaller population leakage. While the physical model of Rydberg blockade gates most often contains simplifications in one way or the other compared to the realistic situations, further optimization in situ can be very helpful while applying the BAM gate idea, including fine-tuning the coefficients of waveforms according to experimentally observed gate fidelities.  

The progress of this work will hopefully reinforce the pursuit of genuinely high connectivity, which may also help the efforts of coherence-preserving mechanical transfer technique and the Rydberg blockade atom-photon gate \cite{OPTICA.5.001492, PRXQuantum.3.010344, PhysRevLett.130.173601}. Apparently, the discussions so far also apply to the situation that the buffer and qubit atoms have different ground-Rydberg transitions, or even they are different elements \cite{PhysRevA.92.042710, PhysRevLett.119.160502, PhysRevLett.128.083202, PhysRevX.12.011040}. The established results can become useful in the crosstalk issue as the distances between qubits can be set at relatively larger values, while the buffer atom potentially provides an extra tool for error correction of the two-qubit system \cite{PhysRevA.106.032425}. Furthermore, the buffer atom can even facilitate the readout process \cite{PhysRevA.92.042710, PhysRevLett.119.180503, PhysRevLett.119.180504, nphys41567, PhysRevApplied.19.054015, PhysRevLett.131.030602, PRJ_Wang22} of qubit atoms via entanglement. Last but not least, for many previously proposed interesting two- and multi- qubit Rydberg blockade gate protocols focusing on two-body interaction \cite{PhysRevA.89.030301, PhysRevA.96.042306, PhysRevA.101.062309, PhysRevX.10.021054, PhysRevLett.127.120501, PhysRevLett.127.120501, PhysRevApplied.16.064031, PhysRevApplied.17.024014, PhysRevA.103.012601, PhysRevApplied.19.044007, ZhangWeiping2023}, the contents of this paper lays down the foundation and establishes the route to upgrade their connectivity via the approach of buffer atom mediated extended two-body interaction, provided such an upgrade is possible, whilst future gate designs may find more and more common ground in the concept and routine of non-trivial modulated driving with carefully calculated waveforms \cite{PhysRevApplied.13.024059, OptEx480513, Yuan2023arXiv}. Overall, it is anticipated that the presented results will serve as a solid building brick of scalability for the cold atom qubit platform.

\begin{acknowledgements}
This work is supported by the National Natural Science Foundation of China (Grant No. 92165107 and No. 12074391) and the fundamental research program of the Chinese Academy of Sciences. The author thanks Ning Chen, Xiaodong He, Zhirong Lin, Tian Xia, Peng Xu and Hui Yan for many discussions.
\end{acknowledgements}

\bibliographystyle{apsrev4-2}

\bibliography{accordion_ref}

\end{document}


\preprint{}

\title{Supplemental Material: \\ Extended two-body Rydberg blockade interaction with off-resonant modulated driving}

\author{Yuan Sun}
\email{yuansun@siom.ac.cn}
\affiliation{CAS Key Laboratory of Quantum Optics and Center of Cold Atom Physics, Shanghai Institute of Optics and Fine Mechanics, Chinese Academy of Sciences, Shanghai 201800, China}

\date{\today}

\maketitle

This supplementary material contains the following contents. (\RN{1}) More details about the atomic transitions of the three-body system formed by the two qubit atoms with one buffer atom. (\RN{2}) Extra discussions about designing waveforms with respect to subtle conditions. More examples of two-qubit entangling gate waveforms will also be provided.

\section{Extra details about the atomic transitions and gate conditions}
\label{sec:part1_details}

In the main text, both the one-photon and two-photon ground-Rydberg transitions have been investigated, and these contents can find straightforward instantiations in alkali and alkali-earth atoms. As widely accepted in experimental efforts of Rydberg atoms, the qubit register states can be chosen as the clock states of alkali or alkali-earth atoms and the two-qubit Rydberg blockade gate can be carried out with the same or different species of atoms \cite{nphys1183, PhysRevLett.104.010503, PhysRevA.92.022336, J.Phys.B.49.202001}. 

\begin{figure}[h]
\centering
\includegraphics[width=0.71\textwidth]{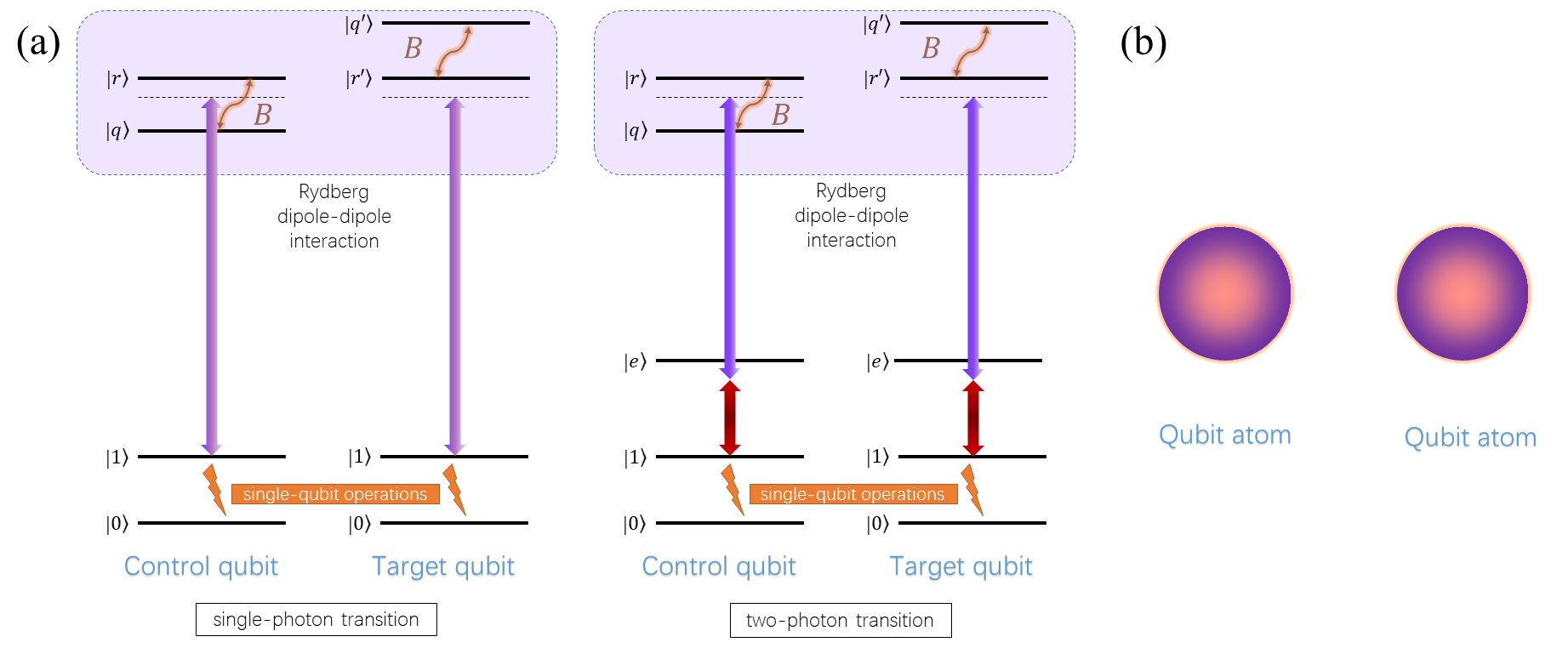}
\caption{(Color online) (a) Typical atomic levels and transitions for two-qubit gate. (b) Cartoon representing the typical geometry of the purely two-body system.}
\label{fig:layout_2body}
\end{figure}

First of all, we revisit the typical two-qubit Rydberg blockade gate with the purely two-body system as shown in Fig. \ref{fig:layout_2body}. For a Rydberg blockade gate with the currently available mainstream experimental hardware, the separation between the two qubit atoms is usually on the order of one to three microns.

\begin{figure}[h]
\centering
\includegraphics[width=0.666\textwidth]{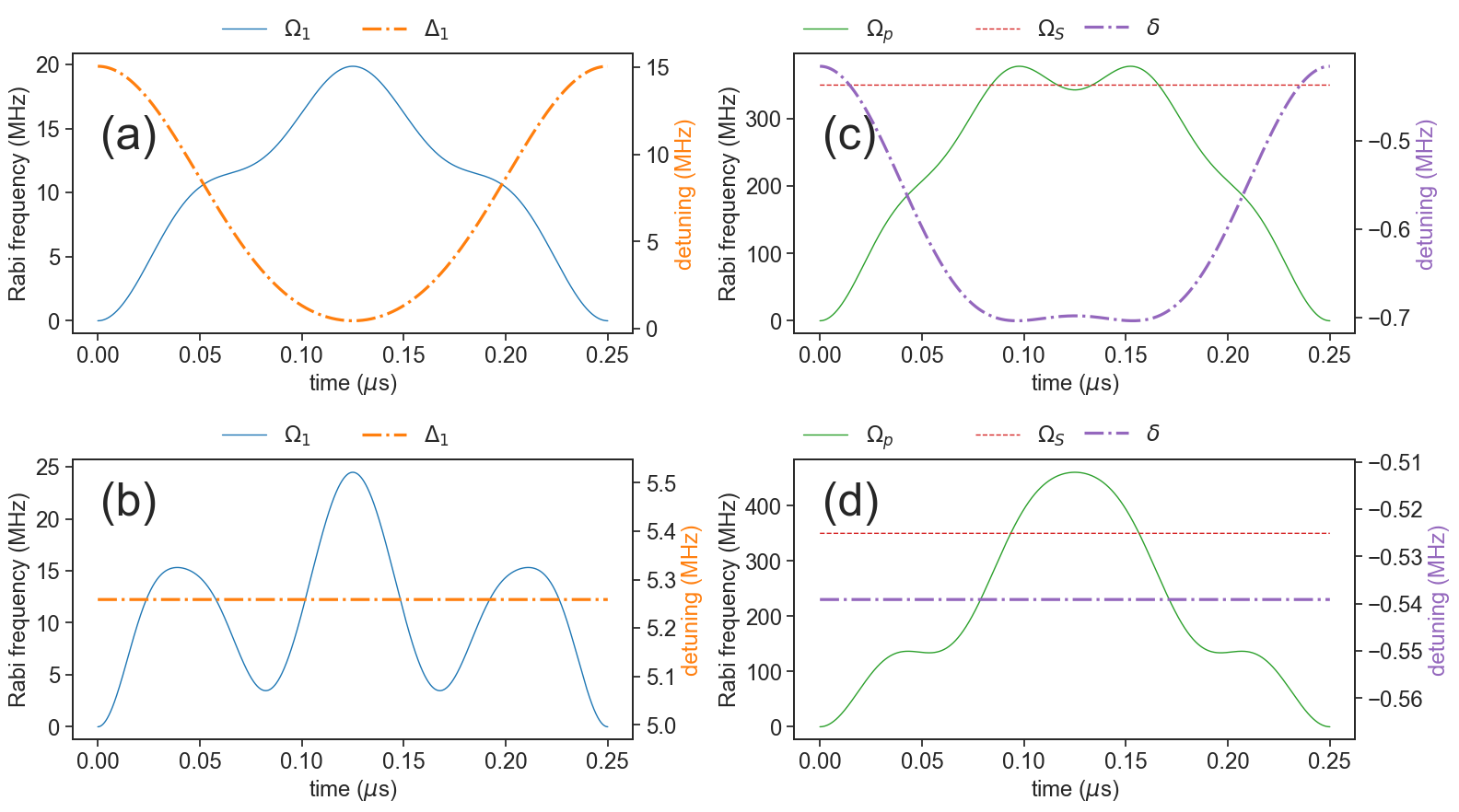}
\caption{(Color online) Two-qubit ORMD gate protocols for close-to-perfect Rydberg blockade effect. (a) One-photon transition, hybrid modulation.  (b) One-photon transition, only amplitude modulation. (c) Two-photon transition, hybrid modulation. (d) Two-photon transition, only amplitude modulation of $\Omega_p$. Here, the condition of homogeneous driving is assumed where both the qubit and control atoms receive the same Rabi frequency and detuning.}
\label{fig:waveforms_v30series}
\end{figure}

In the main text, the concept of tranquilized waveforms is introduced and such a concept certainly can find straightforward applications in the ORMD Rydberg blockade gate of a purely two-body system. ORMD gate protocols designed with this type of waveforms yield analytical pulses that are often friendly to experimentally operate with. In particular, the first-order time derivatives of the starting and ending point of the waveform are zero. Here, typical examples are provided in Fig. S\ref{fig:waveforms_v30series} about applying the tranquilized waveforms to the case of two-qubit Rydberg blockade gate that correspond to the basic ideas of Fig. S\ref{fig:layout_2body}.  

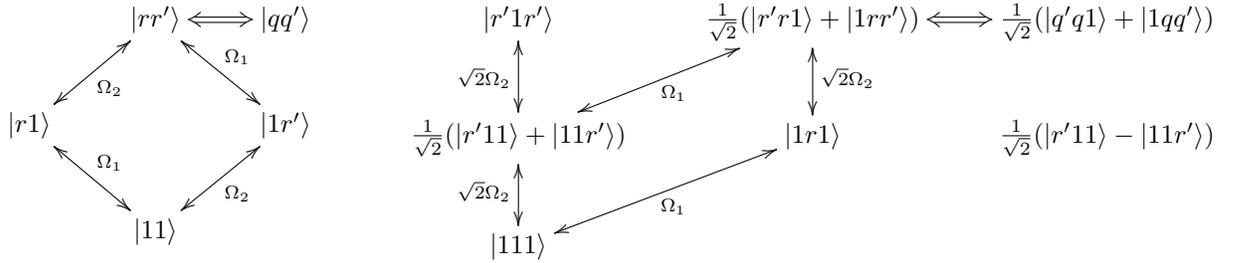
\begin{figure}[h]
\begin{displaymath}
\Scale[1.0]{
\xymatrix{ 
& & |rr'\rangle \ar@{<->}[dl]^{\Omega_2} \ar@{<->}[dr]^{\Omega_1} & |qq'\rangle \ar@{<=>}[l] &  \\
 & |r1\rangle \ar@{<->}[dr]^{\Omega_1} &  & |1r'\rangle \ar@{<->}[dl]^{\Omega_2} & \\ 
 & & |11\rangle & &       
}\,
\xymatrix{
	|r'1r'\rangle \ar@{<->}[d]_{\sqrt{2}\Omega_2} & \frac{1}{\sqrt{2}}(|r'r1\rangle + |1rr'\rangle) \ar@{<->}[d]^{\sqrt{2}\Omega_2} \ar@{<->}[dl]^{\Omega_1}
   & \frac{1}{\sqrt{2}}(|q'q1\rangle + |1qq'\rangle) \ar@{<=>}[l] \\
\frac{1}{\sqrt{2}}(|r'11\rangle + |11r'\rangle)  & |1r1\rangle & 
\frac{1}{\sqrt{2}}(|r'11\rangle - |11r'\rangle) \\
	|111\rangle \ar@{<->}[u]^{\sqrt{2}\Omega_2} \ar@{<->}[ur]_{\Omega_1}  & & 
}			     
            }
\end{displaymath}
\caption{Atomic states and transitions of the three-body system consisting of two qubit and one buffer atoms as discussed in the main text, for one-photon ground-Rydberg transition. The Rabi frequency of buffer atom is denoted as $\Omega_1$ and the Rabi frequency of qubit atom is denoted as $\Omega_2$. The Morris-Shore transform is invoked on the two qubit atoms with identical optical driving for the case of initial state $|111\rangle$. The modeling of Rydberg dipole-dipole interactions is also included in the drawings, as represented by $\Leftrightarrow$.}
\label{fig:1photon_levels}
\end{figure}

Next, we move on the the three-body system of two qubit atoms with one buffer atom, which is the main theme of this work. According to the basic settings of the main text, the two qubit atoms have the same driving fields while the buffer atom has its different driving fields. The relevant atomic states and transition linkage patterns of one-photon ground-Rydberg transition are shown in Fig. S\ref{fig:1photon_levels}. 

The qubit register states are denoted as $|0\rangle, |1\rangle$, where $|0\rangle$ stays out of the ground-Rydberg transition generated by the external driving fields. With the usual choice of rotating wave frame, for the three-body system under study, the purely local dynamics of buffer atom corresponding to the initial state of $|010\rangle$ is described by the Hamiltonian of:
\begin{equation}
\label{eq:Hb_1ph}
H_b = \frac{1}{2}\hbar\Omega_1|1_b\rangle\langle r_b| + \text{H.c.} + \hbar\Delta_1 |r_b\rangle\langle r_b|,
\end{equation}
and the purely local dynamics of qubit atom corresponding to the initial state of $|100\rangle, |001\rangle$ is described by the Hamiltonian of:
\begin{equation}
\label{eq:Hu_1ph}
H_u = \frac{1}{2}\hbar\Omega_2|1_u\rangle\langle r'_u| + \text{H.c.} + \hbar\Delta_2 |r'_u\rangle\langle r'_u|,
\end{equation}
with the subscripts $b, u$ standing for buffer and qubit atoms respectively.
Next, initial states of $|110\rangle$ or $|011\rangle$ involves the two-body dynamics of one buffer and one qubit atoms, whose Hamiltonian can be expressed as:
\begin{align}
\label{eq:Hs_1ph}
H_s &= \frac{1}{2}\hbar\Omega_1|1_b1_u\rangle\langle r_b1_u| + \frac{1}{2}\hbar\Omega_2|1_b1_u\rangle\langle 1_br'_u|
+ \frac{1}{2}\hbar\Omega_2|r_b1_u\rangle\langle r_br'_u| + \frac{1}{2}\hbar\Omega_1|1_br'_u\rangle\langle r_br'_u| + \text{H.c.}\nonumber\\
&+ \hbar\Delta_1|r_b1_u\rangle\langle r_b1_u| + \hbar\Delta_2|1_br'_u\rangle\langle 1_br'_u| + \hbar(\Delta_1+\Delta_2)|r_br'_u\rangle\langle r_br'_u| \nonumber\\
&+ \hbar B|r_br'_u\rangle \langle q_bq'_u| + \hbar B|q_bq'_u\rangle \langle r_br'_u|  + \hbar(\Delta_1+\Delta_2 + \delta_q)|q_bq'_u\rangle\langle q_bq'_u|,
\end{align}
where $B$ is the Rydberg blockade strength and $\delta_q$ is the energy penalty term of state $|qq'\rangle$ in F\"orster resonance. For succinctness, we let $\delta_q = 0$ throughout the text.
Finally, the initial state of $|111\rangle$ engages the system into the three-body interaction, whose Hamiltonian can be written as:
\begin{align}
\label{eq:Hd_1ph}
H_d &= \frac{\sqrt{2}}{2}\hbar\Omega_2|111\rangle\frac{1}{\sqrt{2}}(\langle r'11| + \langle 11r'|) + \frac{1}{2}\hbar\Omega_1|111\rangle\langle 1r1| + \frac{\sqrt{2}}{2}\hbar\Omega_2\frac{1}{\sqrt{2}}(|r'11\rangle+|11r'\rangle)\langle r'1r'| \nonumber\\
&+\frac{1}{2}\hbar\Omega_1\frac{1}{\sqrt{2}}(|r'11\rangle+|11r'\rangle)\frac{1}{\sqrt{2}}(\langle r'r1| + \langle 1rr'|) 
+\frac{\sqrt{2}}{2}\Omega_2|1r1\rangle\frac{1}{\sqrt{2}}(\langle r'r1| + \langle 1rr'|) + \text{H.c.}\nonumber\\
&+ \hbar\Delta_1|1r1\rangle\langle 1r1| + \hbar\Delta_2\frac{1}{\sqrt{2}}(|r'11\rangle+|11r'\rangle)\frac{1}{\sqrt{2}}(\langle r'11| + \langle 11r'|) + 2\hbar\Delta_2|r'1r'\rangle\langle r'1r'|\nonumber\\ 
&+ \hbar(\Delta_1+\Delta_2)\frac{1}{\sqrt{2}}(|r'r1\rangle + | 1rr'\rangle)\frac{1}{\sqrt{2}}(\langle r'r1| + \langle 1rr'|)
+ \hbar(\Delta_1+\Delta_2+\delta_q)\frac{1}{\sqrt{2}}(|q'q1\rangle + | 1qq'\rangle)\frac{1}{\sqrt{2}}(\langle q'q1| + \langle 1qq'|) \nonumber\\
&+ \hbar B \frac{1}{\sqrt{2}}(|r'r1\rangle + | 1rr'\rangle)\frac{1}{\sqrt{2}}(\langle q'q1| + \langle 1qq'|) 
+ \hbar B \frac{1}{\sqrt{2}}(|q'q1\rangle + | 1qq'\rangle)\frac{1}{\sqrt{2}}(\langle r'r1| + \langle 1rr'|),
\end{align}
where the order of labeling states is the same as in the main text, namely that $|111\rangle$ stands for $|1_u1_b1_u\rangle$, where the qubit atoms can be control or target. The linkage patterns of $H_s, H_d$ are shown in Fig. S\ref{fig:1photon_levels}. 

\begin{figure}[h]
\begin{displaymath}
\Scale[0.8]{
\xymatrix{ 
  & & & \\
  & |rr'\rangle \ar@{<=>}[r] \ar@{<->}[dl]^{\Omega_{2S}} \ar@{<->}[dr]^{\Omega_{1S}}  & |qq'\rangle  \\
  |re'\rangle \ar@{<->}[d]^{\Omega_{2p}} &  & |er'\rangle \ar@{<->}[d]^{\Omega_{1p}} \\
  |r1\rangle \ar@{<->}[d]^{\Omega_{1S}} &  & |1r'\rangle \ar@{<->}[d]^{\Omega_{2S}} \\
  |e1\rangle \ar@{<->}[dr]^{\Omega_{1p}} &  & |1e'\rangle \ar@{<->}[dl]^{\Omega_{2p}}  \\ 
  & |11\rangle &    
}\,
\xymatrix{
	|r'er'\rangle \ar@{<->}[d]_{\Omega_{1p}} &  & 
  \frac{1}{\sqrt{2}}(|r're'\rangle + |e'rr'\rangle) \ar@{<->}[d]_{\Omega_{2p}} & \\
  |r'1r'\rangle \ar@{<->}[d]^{\sqrt{2}\Omega_{2S}} & & \frac{1}{\sqrt{2}}(|r'r1\rangle + |1rr'\rangle) \ar@{<->}[d]^{\Omega_{2S}} &
  \frac{1}{\sqrt{2}}(|q'q1\rangle + |1qq'\rangle) \ar@{<=>}[l] \\
  \frac{1}{\sqrt{2}}(|r'1e'\rangle + |e'1r'\rangle) \ar@{<->}[d]_{\Omega_{2p}} & 
  \frac{1}{\sqrt{2}}(|r'e1\rangle + |1er'\rangle) \ar@{<->}[ur]_{\Omega_{1S}} \ar@{<->}[dl]_{\Omega_{1p}}
  & \frac{1}{\sqrt{2}}(|e'r1\rangle + |1re'\rangle) \ar@{<->}[d]_{\sqrt{2}\Omega_{2p}} & 
  \frac{1}{\sqrt{2}}(|r'1e'\rangle - |e'1r'\rangle) \ar@{<->}[d]_{\Omega_{2p}} \\
  \frac{1}{\sqrt{2}}(|r'11\rangle + |11r'\rangle) \ar@{<->}[d]^{\Omega_{2S}} & & |1r1\rangle \ar@{<->}[d]^{\Omega_{1S}} & 
  \frac{1}{\sqrt{2}}(|r'11\rangle - |11r'\rangle) \ar@{<->}[d]^{\Omega_{2S}} \\  
  \frac{1}{\sqrt{2}}(|e'11\rangle + |11e'\rangle) \ar@{<->}[d]_{\sqrt{2}\Omega_{2p}} & & |1e1\rangle \ar@{<->}[dll]_{\Omega_{1p}} & 
  \frac{1}{\sqrt{2}}(|e'11\rangle - |11e'\rangle) \\
  |111\rangle &  &  & 
}			     
            }
\end{displaymath}
\caption{Atomic states and transitions for two-photon ground-Rydberg transition. The Rabi frequencies of buffer atom are denoted as $\Omega_{1p}, \Omega_{1S}$ and the Rabi frequencies of qubit atom are denoted as $\Omega_{2p}, \Omega_{2S}$. The Morris-Shore transform applies to the two qubit atom with identical optical driving for the case of initial state $|111\rangle$. The modeling of Rydberg dipole-dipole interactions is also included in the drawings, as represented by $\Leftrightarrow$. For brevity, the plot of $|11\rangle$ omits the couplings associated with state $|ee'\rangle$, while the plot of $|111\rangle$ omits the couplings of $\frac{1}{\sqrt{2}}(|e'11\rangle + |11e'\rangle) \leftrightarrow \frac{1}{\sqrt{2}}(|e'1e'\rangle + |e'1e'\rangle)$, $\frac{1}{\sqrt{2}}(|e'11\rangle + |11e'\rangle) \leftrightarrow \frac{1}{\sqrt{2}}(|e'e1\rangle + |1ee'\rangle)$ and $\frac{1}{\sqrt{2}}(|e'r1\rangle + |1re'\rangle) \leftrightarrow \frac{1}{\sqrt{2}}(|e're'\rangle + |e're'\rangle)$.}
\label{fig:2photon_levels}
\end{figure}
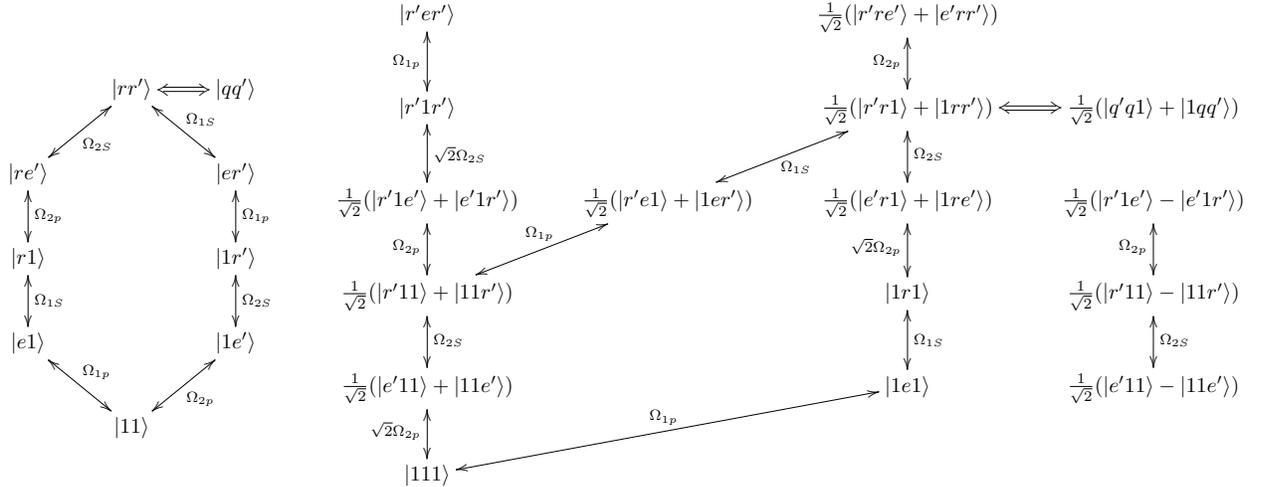

For typical two-photon ground-Rydberg transitions \cite{PhysRevA.105.042430, OptEx480513}, the ground and Rydberg states are coupled via an extra intermediate state with a relatively large one-photon detuning $\Delta_0$. Denote the intermediate states of buffer and qubit atoms as $|e\rangle, |e'\rangle$ respectively, then the Hamiltonian $H_b$ representing the local dynamics of the buffer atom and the Hamiltonian $H_u$ representing the local dynamics of the qubit atoms are:
\begin{equation}
\label{eq:Hb_2ph}
H_b = \frac{1}{2}\hbar\Omega_{1p}|1_b\rangle\langle e_b| + \frac{1}{2}\hbar\Omega_{1S}|e_b\rangle\langle r_b| + \text{H.c.} 
+ \hbar\Delta_0|e_b\rangle\langle e_b| + \hbar\delta_1 |r_b\rangle\langle r_b|,
\end{equation}
\begin{equation}
\label{eq:Hu_2ph}
H_u = \frac{1}{2}\hbar\Omega_{2p}|1_u\rangle\langle e'_u| + \frac{1}{2}\hbar\Omega_{2S}|e'_u\rangle\langle r'_u| + \text{H.c.} + \hbar\Delta_0|e'_u\rangle\langle e'_u| + \hbar\delta_2 |r'_u\rangle\langle r'_u|.
\end{equation}
The detailed states and linkage structures of $H_s, H_d$ for the two-photon ground-Rydberg transition are demonstrated in Fig. S\ref{fig:2photon_levels}. The differences with the single-photon transition versions of Eq. \ref{eq:Hs_1ph} and Eq. \ref{eq:Hd_1ph} are the replacement of two-photon ground-Rydberg couplings as in Eq. \ref{eq:Hb_2ph} and Eq. \ref{eq:Hu_2ph}.

The two-photon ground-Rydberg transition under discussion satisfies the prerequisite of adiabatic elimination. Namely, the one-photon detuning $\Delta_0$ is relatively large compared with the Rabi frequency terms $\Omega_p, \Omega_S$. Moreover, choosing the appropriate intermediate states needs some careful evaluations, including the requirement of line-width $\lesssim 1$ MHz for $\sim 100$ ns gate time, in order to suppress gate errors on the order of $1 \times 10^{-4}$ in the future.

So far, we have kept the complexities of atomic states and coherent transitions to a relatively minimal extent in the description of the system. This suffices for the purpose of this work, as these discussions are reasonable and adequate to represent the underlying physics. On the other hand, it can be perceived as an over-simplification of the actual atom-light interaction processes if compared to a serious experimental study, which involve more states and transitions. In fact, these extra states and transitions can be accurately analyzed with respect to practical situations if necessary, which does not impede the main results of this work.

\begin{figure}[h]
\centering
\includegraphics[width=0.666\textwidth]{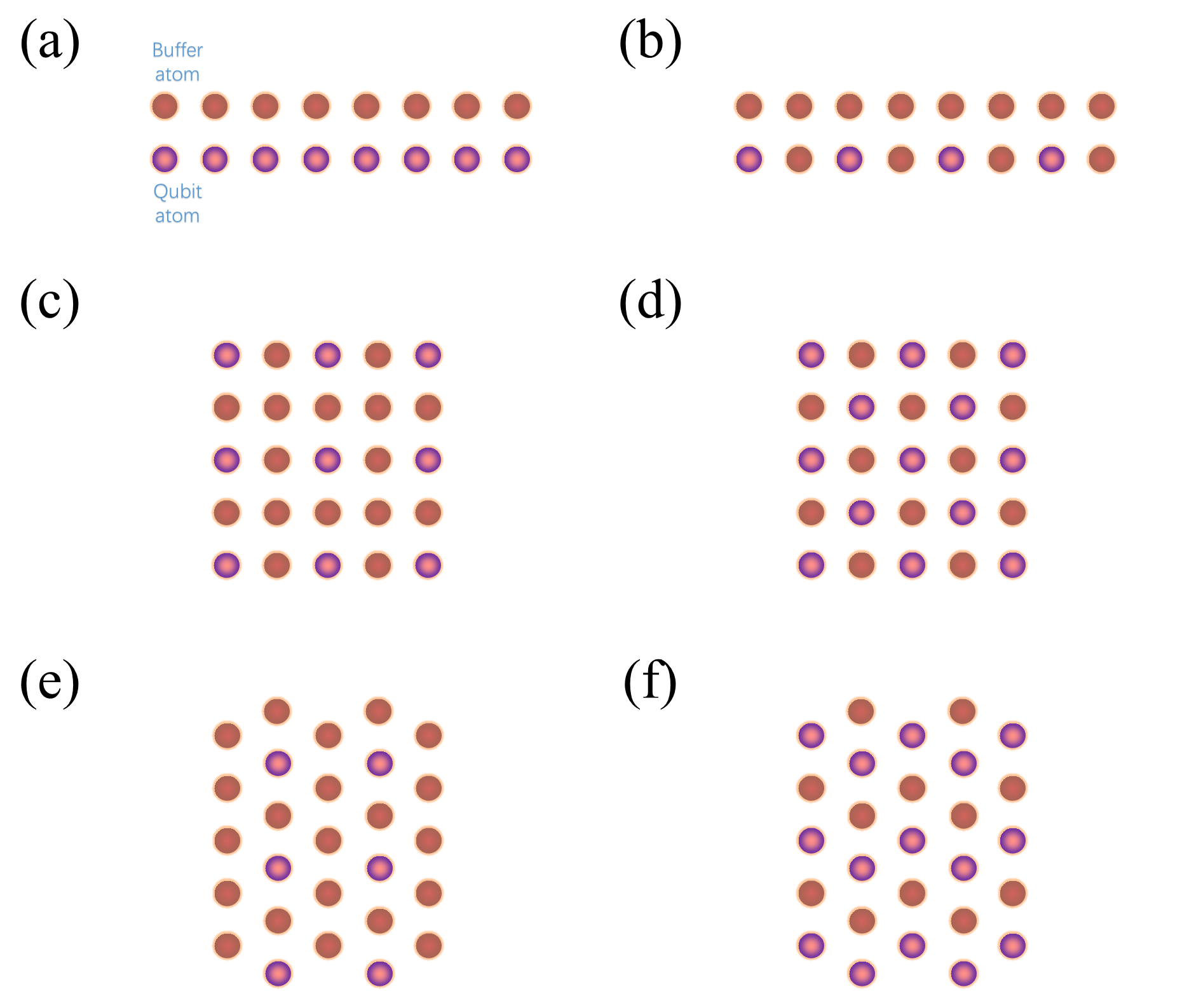}
\caption{Sketch of cold atom qubit array built with the buffer atom relay.}
\label{fig:buffer_array_framework}
\end{figure}

The main text discusses the idea of buffer atom relay to connect two very far away qubit atoms. This idea provides a potential solution to systematically construct a fully-connected cold atom qubit array. It is anticipated the arrangement of qubit atoms and buffer atom relays can also help in other aspects such as atom-photon gates and coupling to solid-state devices. Fig. S\ref{fig:buffer_array_framework} shows some typical geometric configurations conforming to the concept of cold atom qubit array with the buffer atom relay.

\section{Extra examples of BAM gates with ORMD method}
\label{sec:part2_nuances}

\begin{figure}[h]
\centering
\includegraphics[width=0.666\textwidth]{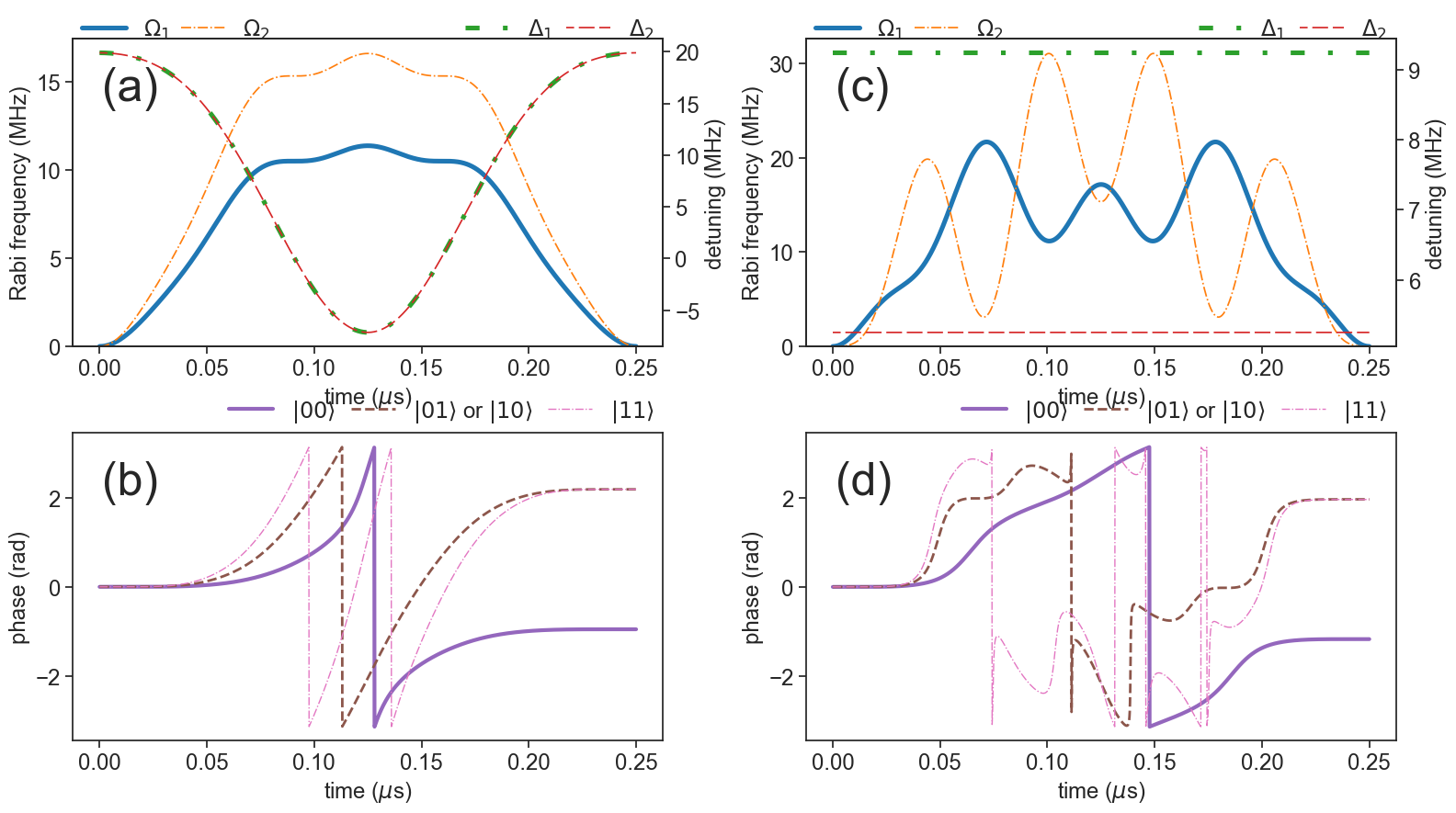}
\caption{Sample waveforms of BAM gate with one buffer atom via one-photon ground-Rydberg transition. $B=2\pi\times100\text{ MHz}, \delta_q=0$. (a) Hybrid modulation.  (b) Phases of wave functions corresponding to (a). (c) Only amplitude modulation.  (d) Phases of wave functions corresponding to (c). The calculated gate errors are less than $10^{-4}$.}
\label{fig:B100MHz_1ph}
\end{figure}

\begin{figure}[h]
\centering
\includegraphics[width=0.666\textwidth]{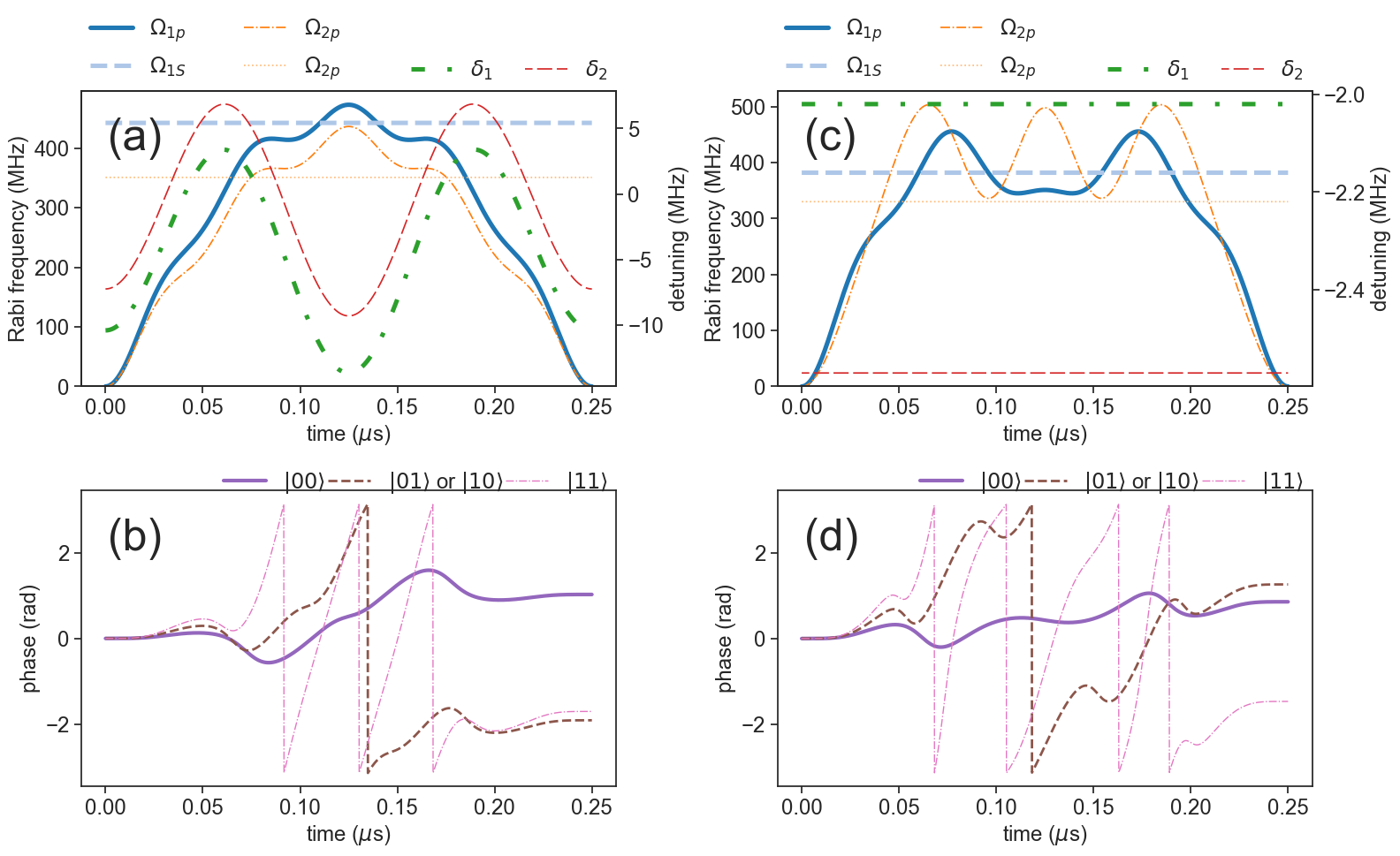}
\caption{Sample waveforms of BAM gate with one buffer atom via two-photon ground-Rydberg transition. $B=2\pi\times100\text{ MHz}, \delta_q=0$. (a) Hybrid modulation. (b) Phases of wave functions corresponding to (a). (c) Only amplitude modulation.  (d) Phases of wave functions corresponding to (c). The calculated gate errors are less than $10^{-4}$.}
\label{fig:B100MHz_2ph}
\end{figure}

It certainly makes sense to tailor Rydberg blockade gates according to the experimentally measured value of Rydberg blockade strength instead of using the strong assumption of perfect Rydberg blockade. In the main text, the calculations of waveforms all stick to the condition of $B=2\pi\times50\text{ MHz}, \delta_q=0$ as a specific example. Obviously, the BAM gate method can adapt to various different values without much trouble, following the routine discussed in the main text.  

A few extra examples are provided here with respect to $B=100\text{ MHz}, \delta_q=0$. Fig. S\ref{fig:B100MHz_1ph} demonstrates BAM gate waveforms with respect to the one-photon ground-Rydberg transition. For Fig. S\ref{fig:B100MHz_1ph}(a), $\Omega_1(t) = 0.685\times\Omega_2(t)$, $\Omega_2(t)$ is given by [111.45, -44.05, -11.36, 0.33, 1.33, -1.97], $\Delta_1(t)$ is given by [42.79, 33.82, -5.44] and $\Delta_2(t)$ is given by [42.79, 33.82, -5.44]. For Fig. S\ref{fig:B100MHz_1ph}(c), $\Omega_1(t), \Omega_2(t)$ are expressed by [128.31, -35.90, -29.34, 1.79, 12.40, -13.11],  [150.75, -50.32, 2.47, -24.40, -35.65, 32.53] respectively, $\Delta_1=2\pi\times9.24\text{ MHz}$ and $\Delta_2=2\pi\times5.25 \text{ MHz}$. Fig. S\ref{fig:B100MHz_2ph} demonstrates BAM gate waveforms with respect to the two-photon ground-Rydberg transition, and the one-photon detuning is the same as the main text: $\Delta_0=2\pi\times5\text{ GHz}$. For Fig. S\ref{fig:B100MHz_2ph}(a), $\Omega_{1S}=2\pi\times442.41\text{ MHz}, \Omega_{2S}=2\pi\times350.90\text{ MHz}$, and $\Omega_{1p}(t), \Omega_{2p}(t), \delta_1(t), \delta_2(t)$ are given by [3234.43, -1086.98, -275.00, -93.96, -41.85, -119.42], [2850.37, -995.04, -196.13, -84.17, -28.95, -120.89], [-21.68, 4.06, -19.20] and [-3.57, 2.57, -18.87] respectively. For Fig. S\ref{fig:B100MHz_2ph}(c), $\Omega_{1p}(t), \Omega_{2p}(t)$ are given by [3337.20, -767.85, -614.38, -41.25, -88.80, -156.33], [3624.89, -894.35, -690.53, -346.77, 247.19, -127.99] respectively, and $\Omega_{1S}=2\pi\times381.88\text{ MHz}, \Omega_{2S}=2\pi\times330.08\text{ MHz}, \delta_1=2\pi\times-2.02\text{ MHz}, \delta_2=2\pi\times-2.57\text{ MHz}$.

\begin{figure}[h]
\centering
\includegraphics[width=0.666\textwidth]{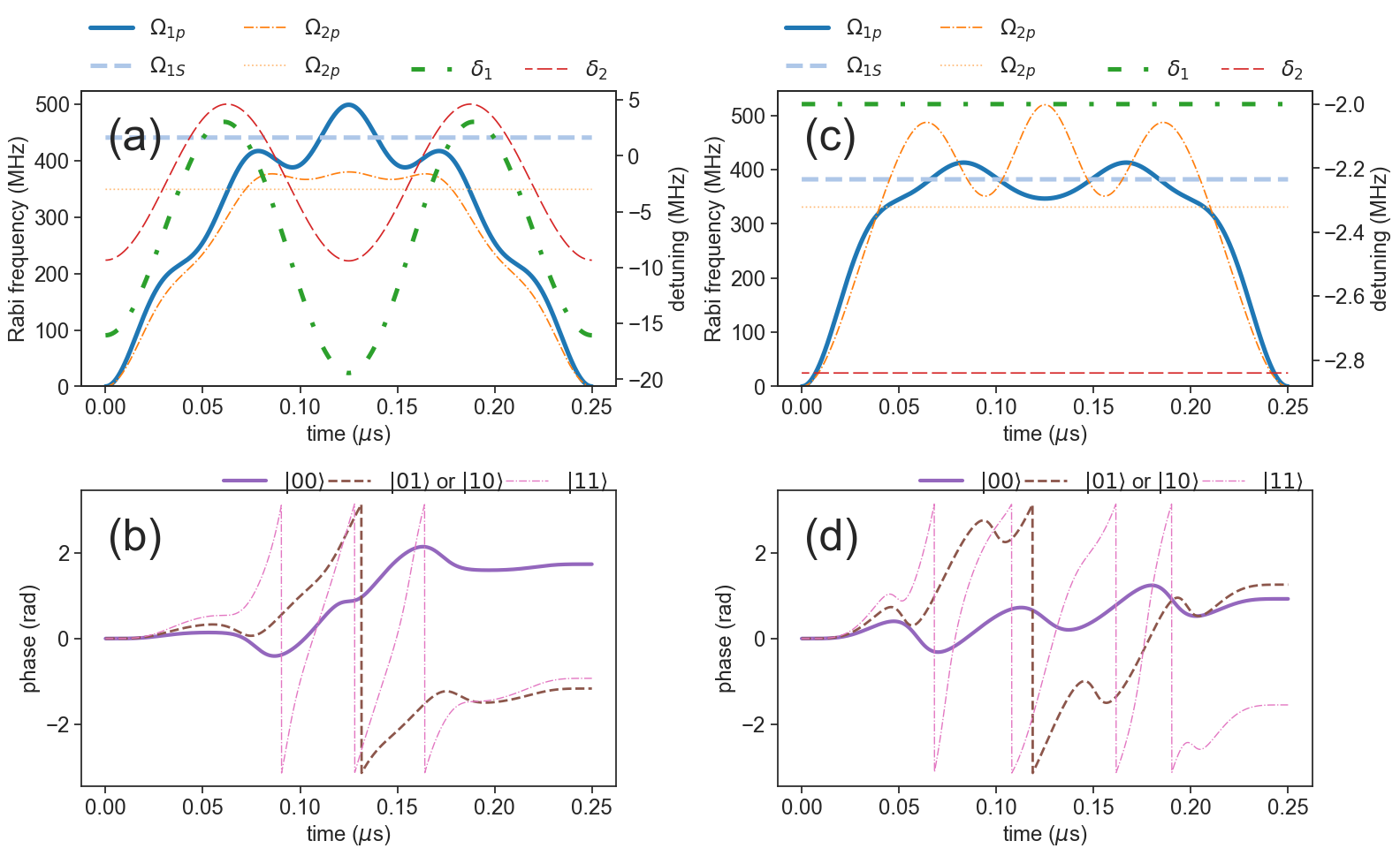}
\caption{Examples of BAM gate waveforms for case of non-negligible Rydberg dipole-dipole interaction strengths between the two qubit atoms, via two-photon ground-Rydberg transition. (a) Hybrid modulation. (b) Phases of wave functions corresponding to (a). (c) Only amplitude modulation.  (d) Phases of wave functions corresponding to (c). The calculated gate errors are less than $10^{-4}$.}
\label{fig:B100MHz_2ph_consider_risa}
\end{figure}

About three-body system formed by one buffer atom and two qubit atoms, the derivations in the main text assume that the Rydberg dipole-dipole interaction of two faraway qubit atoms is comparatively weak or very close to zero under certain circumstances, and therefore neglect it for the moment. On the other hand, with respect to possible experimental situations that the two qubit atoms have non-negligible long-range Rydberg dipole-dipole interactions, higher level of accuracy can be obtained in the gate design process if the residual Rydberg dipole-dipole interaction is included into consideration. In fact, we expect that high-fidelity results can still be obtained with relatively small changes on the waveforms which have neglected this subtle effect. As a specific example, we postulate that the buffer and qubit atoms are evenly spaced, the target Rydberg states are the same for all the buffer and qubit atoms, and the scaling law of Rydberg dipole-dipole interaction strength obeys $1/R^6$ with respect to inter-atomic distance $R$. In other words, for the previously discussed situations as depicted in Fig. S\ref{fig:1photon_levels} and S\ref{fig:2photon_levels} where $B=2\pi\times100\text{ MHz}, \delta_q=0$ for the Rydberg dipole-dipole interaction between one buffer and one qubit atom, an extra energy shift of $\delta_r=2\pi\times100/64\text{ MHz}=2\pi\times1.5625\text{ MHz}$ is introduced between the two qubit atoms in the state of $|r'\rangle$. The BAM gate waveforms of this specific example are demonstrated in Fig. S\ref{fig:B100MHz_2ph_consider_risa}. For Fig. S\ref{fig:B100MHz_2ph_consider_risa}(a), $\Omega_{1S}=2\pi\times440.79\text{ MHz}, \Omega_{2S}=2\pi\times349.13\text{ MHz}$, and $\Omega_{1p}(t), \Omega_{2p}(t), \delta_1(t), \delta_2(t)$ are given by [3244.11, -1061.56, -262.13, -112.21, 12.12, -198.28], [2825.37, -922.56, -302.26, -34.04, -65.55, -88.27], [-36.85, 4.22, -25.95] and [-11.85, 0.08, -17.50] respectively. For Fig. S\ref{fig:B100MHz_2ph_consider_risa}(c), $\Omega_{1p}(t), \Omega_{2p}(t)$ are given by [3348.93, -726.20, -546.18, -159.50, -174.75, -67.84], [3671.33, -936.72, -622.74, -393.10, 216.80, -99.90] respectively, and $\Omega_{1S}=2\pi\times382.25\text{ MHz}, \Omega_{2S}=2\pi\times331.04\text{ MHz}, \delta_1=2\pi\times-2.00\text{ MHz}, \delta_2=2\pi\times-2.84\text{ MHz}$.


\bibliography{accordion_ref}